\documentclass[twocolumn,preprintnumbers,amsmath,amssymb]{revtex4}
\usepackage{graphicx}
\catcode`\@=11
\def\eqlt{\mathrel{\mathpalette\@vereq<}}  
\def\eqgt{\mathrel{\mathpalette\@vereq>}}  
\def\@vereq#1#2{\lower2.5pt\vbox{\baselineskip0pt \lineskip-.5pt
 \ialign{$\m@th#1\hfil##\hfil$\crcr#2\crcr{=}\crcr}}}

\newcommand{\simle}{\ \raise.3ex\hbox{$<$}\kern-0.8em\lower.7ex\hbox{$\sim$}\ }
\newcommand{\simge}{\ \raise.3ex\hbox{$>$}\kern-0.8em\lower.7ex\hbox{$\sim$}\ }

\begin{document}
\title {Electronic Structure of Strongly Correlated Systems Emerging from Combining Path-Integral Renormalization Group with Density Functional Approach}  
\author {Yoshiki Imai, Igor V. Solovyev and Masatoshi  Imada}  
\address {Institute for Solid State Physics, University of Tokyo, Kashiwanoha, Kashiwa, Chiba, 277-8581, Japan and \\ 
PRESTO, Japan Science and Technology Agency}  

\begin{abstract} 
A new scheme of first-principles computation for strongly correlated electron systems is proposed.  This scheme starts from the local-density approximation (LDA) at high-energy band structure, while the low-energy effective Hamiltonian is constructed by a downfolding procedure using combinations of the constrained LDA and the GW method.  Thus obtained low-energy Hamiltonian is solved by the path-integral renormalization-group method, where spatial and dynamical fluctuations are fully considered.  An application to Sr$_2$VO$_4$ shows that the scheme is powerful in agreement with experimental results. It further predicts a nontrivial orbital-stripe order.     
\end{abstract}
\maketitle
First principles methods for electronic structure calculations have been extensively applied to various systems in the last few decades.  The density functional theory (DFT) supplemented with the local-density-approximation (LDA)~\cite{DFT,DFT2} is one of the most successful schemes.
However, in strongly correlated electron systems such as transition metal oxides and organic compounds, LDA fails even at qualitative levels of simple systems and remain challenges in terms of first-principles calculations.  The failure is ascribed to the correlation effects generating large spatial and dynamical fluctuations beyond the applicability of existing schemes.
  

In electronic structure calculations, DFT and wavefunction methods are two typical approaches~\cite{Kohn}. The rapid development and recent extensive applications of DFT approach rely on its less computation time while a systematic improvement of its accuracy is not an easy task.  On the other hand, the wavefunction methods including the Hartree-Fock and variational schemes cost much longer computational time, while it offers a better accuracy particularly for strongly correlated systems as we will discuss later.  

After considering this contrast, an optimal choice of the algorithm would be a hybrid method. Since the electrons far from the Fermi level do not show serious fluctuation effects, LDA and GW~\cite{GW1,GW2} offer a reasonable and efficient computational scheme there. If the tracing out and the elimination of the high-energy degrees of freedom by a downfolding scheme would result in the effective low-energy Lagrangian or Hamiltonian, they can be treated by an accurate wavefunction method. Such successive eliminations of the high-energy part have a conceptual similarity to the renormalization group method. In this paper, we propose a downfolding scheme to derive the low-energy effective models from the higher-energy structure, together with a reliable low-energy solver for the downfolded effective model, as a single framework of a first-principles method.  This offers a powerful first-principle method for strongly correlated electron systems.   To show the performance, we apply the present scheme to Sr$_2$VO$_4$, which reveals intriguing properties of this compound.

Since the electrons have kinetic and interaction energies, the downfolding procedure basically consists of two parts. One is to derive the screened Coulomb interaction for low-energy electrons after eliminating the high-energy degrees of freedom. The other is to take into account the effect on the single-particle kinetic part, which can be expressed as the self-energy effect from the high-energy electrons, if the quasi-particle description is justified.  

After eliminating the high-energy degrees of freedom, the effective model in general has a frequency dependence represented by a Lagrangian.  However, in the low-energy region, the Hamiltonian approach by replacing the dynamical Coulomb interaction with the static one still offers an efficient and essentially correct framework if a certain condition is satisfied~\cite{Aryasetiawan,Solovyev}.
        
Recently, the dynamical mean-field theory (DMFT)~\cite{DMFT} was employed as a low-energy solver for finite temperature properties~\cite{DMFT+GW}.  Since the DMFT ignores the spatial correlations, its refinements by using the cluster or cellular algorithms~\cite{Cellular,Cluster} were attempted.
In this paper, we alternatively solve the low-energy effective Hamiltonian by the path-integral renormalization group (PIRG) method~\cite{ImadaKashima} implemented with quantum number projection algorithm~\cite{Mizusaki}. 
The PIRG scheme allows treating equally the spatial and dynamical fluctuations in a controllable way and hence makes it possible to obtain the ground state of the effective Hamiltonian exactly in a numerical optimization scheme.   

We now start with the downfolding procedure.
We first perform the LDA calculations based on the LMTO basis functions~\cite{LMTO}.  
In many correlated electron systems, the bands close to the Fermi level are relatively well separated from the high-energy band structure as in transition metal oxides, where the bands near the Fermi level consist of 3$d$ character mainly derived from transition-metal atomic orbitals (and frequently hybridized with oxygen 2$p$ atomic orbitals). In this circumstance, the whole bands may be downfolded to an effective model consisting only of the transition metal $3d$ bands (or additionally with the oxygen $2p$ bands).  If the crystal field splitting is strong, it may be downfolded even to a set of the bands closest to the Fermi level, for example, the $t_{2g}$ or $e_g$ bands only, in a cubic environment.        
Here we report the first attempt of the downfolding along this line.  

\begin{figure}
\includegraphics[width=8.5cm]{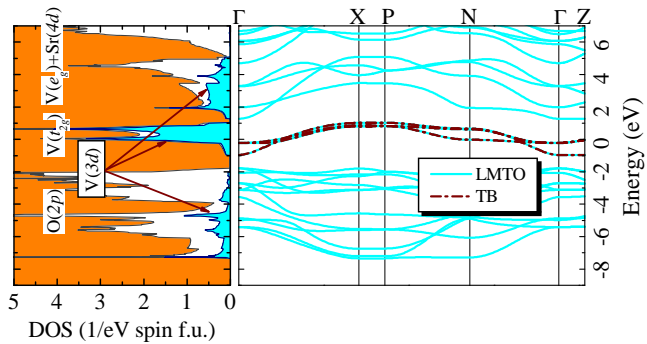}
\caption{(color) (Right panel):Comparison of LDA dispersions for Sr$_2$VO$_4$ computed from the LMTO basis functions (dotted curve) with the downfolded tight-binding bands (dot-dashed curve). (Left panel):The corresponding LDA density of states. }
\label{Fig.SVOLDADispersion}
\end{figure}
When we restrict the degrees of freedom to the isolated LDA bands closest to the Fermi level, we first need to reduce the Hilbert space. 
With the notation for the states having the maximal weight near the Fermi level as
$\{ | d \rangle \}$ and the rest of the basis functions as $\{ | r \rangle \}$,
the whole Hilbert space may be spanned as $\{ | \chi \rangle \}$$=$$\{ | d \rangle \}$$\oplus$$\{ | r \rangle \}$. A typical example of Sr$_2$VO$_4$ is shown in Fig.~\ref{Fig.SVOLDADispersion}. In this case, three isolated bands closest to the Fermi level mainly consist of the $3d$ $t_{2g}$ states, which defines the Hilbert subspace $\{ |d\rangle\}$ in our downfolding procedure.
The LDA eigenvalue equation can be rewritten as
\begin{eqnarray}
( \hat{H}^{dd}-\omega ) | d \rangle  +  \hat{H}^{dr} | r \rangle & = & 0, \label{eqn:seceq1}\\
\hat{H}^{rd} | d \rangle  +  ( \hat{H}^{rr}-\omega ) | r \rangle & = & 0. \label{eqn:seceq2}
\end{eqnarray}
We eliminate the subspace $| r \rangle$, 
which yields the effective
$\omega$-dependent Hamiltonian for only $d$-subspace:
$
\hat{H}^{dd}_{\rm eff}(\omega) = \hat{H}^{dd} - \hat{H}^{dr}
(\hat{H}^{rr} - \omega)^{-1}\hat{H}^{rd}.
$
From the overlap matrix
$
\hat{S}(\omega)=1+\hat{H}^{dr}
(\hat{H}^{rr}-\omega)^{-2}\hat{H}^{rd},
$
with $\langle d | \hat{S} | d \rangle$$=$$1$,
the tight-binding effective Hamiltonian, $\hat{h}$ is obtained after the orthonormalization
of the vectors $| d \rangle$$\rightarrow$$| \tilde{d} \rangle$$=$$\widehat{S}^{1/2}| d \rangle$
and fixing the energy $\omega$ in the center of gravity of the band of our interest:
$
\hat{h} = \hat{S}^{-1/2} \hat{H}^{dd}_{\rm eff} \hat{S}^{-1/2}.
$
After Fourier transformation to the real space, $\hat{h}$ defines the {\it Wannier basis}. 
The band structure obtained after the elimination of the higher-energy bands shows an excellent agreement with the LDA band computed from the LMTO basis as we see in Fig.~\ref{Fig.SVOLDADispersion}. 
This formalism has generality and can be easily extended to a more complex band structure~\cite{PRB04}.

We now derive an effective Coulomb interaction $\hat{W}_r$ among electrons at the isolated $t_{2g}$ Wannier orbitals after considering the screening by other bands.  In principle, $\hat{W}_{r}$ may be derived by a full GW scheme~\cite{Aryasetiawan}.  However, for practical use, the whole GW calculation requires large computation time.  Since the screening from the polarizations of high-energy electrons does not have strong fluctuations, it may alternatively be replaced by the constrained-LDA (C-LDA) method following the procedure in the literature~\cite{UfromconstraintLSDA}. Then we take two steps in deriving the screening.  First, we compute the interaction $\hat{W}_{r1}$ which takes into account the screening by the electrons residing at {\it atomic orbitals} other than $3d$.  This step is performed by the C-LDA method~\cite{UfromconstraintLSDA}.  
When the higher-energy bands are well separated, we expect that the frequency dependence is small around the Fermi level and it may well be calculated by the C-LDA scheme~\cite{UfromconstraintLSDA} by starting with the basic definition of the effective Coulomb interaction formulated by Herring~\cite{Herring}:
$
U_{{\bf RR}'} = E [ n_{{\bf R}}+1,n_{{\bf R}'}-1 ] -
E [ n_{{\bf R}},n_{{\bf R}'}  ],
$
which is nothing but the
energy cost for moving a $3d$-electron between two atoms, located at ${\bf R}$
and ${\bf R}'$, and initially populated by $n_{\bf R}$$=$$n_{{\bf R}'}$$\equiv$$n$
electrons.
Note that at this stage, the screenings of $3d$-$3d$ interactions by the electrons at the same $3d$ {\it atomic} orbitals are excluded, because the $3d$ atomic orbitals do not hybridize each other in C-LDA scheme.
For Sr$_2$VO$_4$, the onsite Coulomb interations derived from $U_{\bf RR'}$ is
$U$$\simeq$$11.3$ eV. For comparison, the bare Coulomb interaction is 21.8 eV.
Thus, the screening of $3d$-$3d$ interactions by non-$3d$ electrons is strong.
In addition to the parameter $U$, the intra-atomic exchange coupling $J$ can be easily calculated in the
framework of C-LDA approach. By using these two parameters, one can restore the full matrix $\hat{W}_{r1}$ of screened Coulomb interactions\cite{PRB94}.

It is known that LDA does not take into account the self-energy effect and fails in reproducing a gap structure at the Fermi level~\cite{GW2}.  LDA also ignores energy shifts in core electron levels arising from the self-interaction effect.  However, in the present scheme, the bands near the Fermi level are left for a more refined PIRG method, while the core electrons hardly polarize and does not contribute to the screening.  The valence electrons give the major contribution to the screening in this C-LDA treatment and their energies are well described by LDA with few self-energy corrections in many cases including transition metal oxides~\cite{Yamazaki-Fujiwara}.  Therefore, although C-LDA is not complete, in the practical usage, it offers a reasonable and efficient way of counting the screening effects by the valence electrons. 

\begin{figure}
\includegraphics[width=8.5cm]{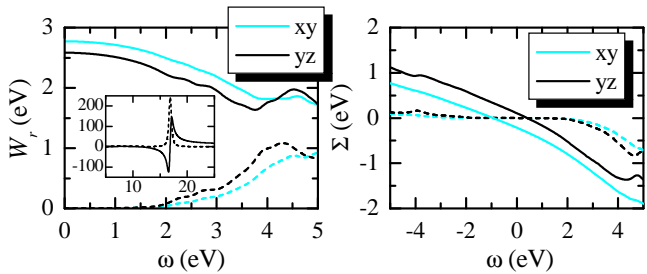}
\caption{(color) Diagonal matrix elements of the screened Coulomb interaction $W_r(\omega)$ (left panel) and the self-energy, $\Sigma(\omega)$ (right panel). The inset shows the high-frequency part of $W_r(\omega)$ (the lines corresponding to the $xy$ and $yz$ orbitals are indistiguishably close). The real and imaginary parts are shown by solid and dashed curves, respectively.}
\label{Fig.SVOsigma}
\end{figure}

In the second step, we use the GW scheme~\cite{Aryasetiawan} by taking $\hat{W}_{r1}$ as if it is the starting bare Coulomb interaction.  Then the RPA screening produces $\hat{W}_{r}(\omega)=\hat{W}_{r1}/(1-\hat{P}_{dr}\hat{W}_{r1})$. The whole polarization of the $3d$ {\it atomic orbitals} in the LMTO basis is given by $\hat{P}_d=\hat{P}_{dr}+\hat{P}_{t2g}$, where $\hat{P}_{t2g}$ is the polarization purely from the $t_{2g}$ {\it LDA band}, while $\hat{P}_{dr}$ represents the rest of $3d$ {\it atomic orbitals} contribution contained in the $e_g$ {\it LDA band} as well as in the component hybridized in the oxygen $2p$ {\it LDA band}.  Since the identity $\hat{W}(\omega) =\hat{W}_{r1}/(1-\hat{P}_{d}\hat{W}_{r1}) =\hat{W}_{r}/(1-\hat{P}_{t2g}\hat{W}_{r})$ holds, $\hat{W}_{r}$ plays the role of the effective interaction in the downfolded Hilbert space of the $t_{2g}$ {\it LDA band}. We also note that some of the oxygen $2p$ LMTO component is contained in the $3d$ $t_{2g}$ {\it Wannier orbital}.  The screening by these components is already taken into account by the C-LDA scheme, while it should be excluded in the low-energy model for the $3d$ $t_{2g}$ {\it Wannier orbitals}. In practice, however, this contribution to the screening turns out to be negligible.  Although the whole GW scheme considering $P_{t2g}$ results in $W$, we consider the interaction within the $t_{2g}$ LDA band more accurately by the low-energy solver as we explained above.  Then in the effective low-energy model for the basis represention of the $t_{2g}$ {\it Wannier orbital}, the   screened Coulomb interaction is given by $\hat{W}_r$.  In general, $\hat{W}_r$ becomes frequency dependent(Fig.~\ref{Fig.SVOsigma}).  If the frequency dependence is small within the range of the $t_{2g}$ bandwidth, we are allowed to take the low-frequency limit $\hat{W}_{r}(\omega=0)$ as the interaction part $\hat{U}$ in the low-energy effective Hamiltonian~\cite{Aryasetiawan}.   

In the downfolding process, the kinetic-energy part is modified through the self-energy $\Sigma(k,\omega)$, which can be evaluated in the GW approximation~\cite{Aryasetiawan}. Here the self-energy effect  from $3d$ and $2p$ atomic orbitals is considered while the polarization of the higher energy bands is small and we neglect it. We also take into account the self-energy arising from the dynamical part of the Coulomb interaction $\hat{W}_{r}(\omega)-\hat{W}_{r}(\omega=0)$ through the GW scheme~\cite{Aryasetiawan}.  Such a self-energy effect coming from the higher energy part of $W_r$ mainly appears through ${\rm Re} \Sigma$, and contributes to the renormalization factor $Z=(1-\partial \Sigma/\partial \omega)^{-1}_{\omega=0} (\sim 0.8$ for Sr$_2$VO$_4$).  
The low-energy part of ${\rm Im} \Sigma$ is small and can be ignored (Fig.\ref{Fig.SVOsigma}).   
Then the effective Hamiltonian is reduced to a multi-band Hubbard model in the {\it Wannier representation} for the $t_{2g}$ band:
\begin{equation}
{\cal H}=\sum_{\stackrel{\scriptstyle \langle i,j\rangle}{m,m',\sigma}}
t^{mm'}_{ij}c^{\dag}_{im\sigma}c_{jm'\sigma}
+\frac{1}{2}\sum_{\stackrel{\scriptstyle i,\alpha,\beta}{\gamma,\delta}}
{U}_{\alpha\beta\gamma\delta} 
c^{\dag}_{i\alpha}c^{\dag}_{i\beta}
 c_{i\gamma}c_{i\delta},
\label{eqn_effham}
\end{equation}
where $c^{\dag}_{im\sigma}(c_{im\sigma})$ creates (annihilates) an electron with the spin $\sigma =(\uparrow, \downarrow)$ at the $t_{2g}$ Wannier orbital $m=(xy,yz,zx)$ of the site $i$ and $n_{im\sigma}=c^{\dag}_{im\sigma}c_{im\sigma}$.  The Greek symbols stand for the combination $(m,\sigma)$ of the indices.  

We now discuss the PIRG method as a solver of the low-energy effective Hamiltonian.
The basic method~\cite{ImadaKashima} is to perform the path-integral operation by following the principle that $\lim_{p\rightarrow \infty}[\exp(-\tau {\cal H})]^p|\Phi\rangle$ generates the ground state, with a proper starting trial wavefunction $|\Phi\rangle$. Through the summation over the Stratonovich variables in the Stratonovich-Hubbard transformation of the interaction term, the obtained state is given by a linear combination of increased number of the nonorthogonal Slater-determinant basis functions.  The optimized basis construction with systematic extrapolation to the full Hilbert space leads to an accurate estimate of the ground state.  It is also crucial to implement the quantum number projection in restoring the symmetry of the Hamiltonian and also in improving the accuracy~\cite{Mizusaki}.  A typical accuracy is seen in the ground-state energy estimate of $-1.85790\pm0.00002$ in comparison with the quantum Monte Carlo result of $-1.8574\pm 0.0014$ for the standard Hubbard model at half filling on a $6\times 6$ square lattice with the nearest neighbor transfer $t=1$ at the onsite interaction $U=4$~\cite{Mizusaki}.  

The present first-principles method combined with the PIRG solver is now applied to calculate the electronic structure of Sr$_2$VO$_4$.
This compound has a layered perovskite structure and bears a intriguing character because this has one 3$d$ electron per V site ($d^1$ system) with strong two-dimensional anisotropy and has a dual relation to the one 3$d$ hole per Cu sites ($d^9$ system) in the mother compounds of the cuprate superconductors.  The duality is not complete because, in Sr$_2$VO$_4$, the orbital degeneracy of $d^1$ electron perfectly remains between $d_{yz}$ and $d_{zx}$ orbitals.  The crystal field splitting of $d_{xy}$ orbital is also rather small ($\sim 0.08$ eV in the LDA calculation).  

In case of Sr$_2$VO$_4$, after the downfolding, the onsite interactions among the Wannier orbitals of intraorbital $xy, yz(zx)$ and interorbital $xy$-$yz(xy$-$zx)$ and $yz$-$zx$ combinations are $U=2.77, 2.58, 1.35$ and 1.28 eV, respectively. The onsite exchange interactions between $xy$-$yz(xy$-$zx)$ and $yz$-$zx$ orbitals are 0.65 and 0.64, respectively.  The transfer between the nearest neighbor $xy$-$xy, yz$-$yz$ and $zx$-$zx$ orbitals in $x$ direction are -0.22, -0.05 and -0.19 eV, respectively. Although the transfer between the next-nearest-neighbor pairs are considerably smaller, they have also been taken into account in our calculations.  
In order to monitor the effect of Coulomb interaction, we examine the scale-factor dependence by multiplying all the matrix elements $U_{\alpha\beta\gamma\delta}$ in Eq.(\ref{eqn_effham}
) with a factor $\lambda$.  Namely, the realistic value in the downfolded Hamiltonian corresponds to $\lambda=1$.

Sr$_2$VO$_4$ was first experimentally studied by Cyrot {\it et al.} and Nozaki {\it et al.}~\cite{Cyrot,ISTEC} and recently thin film formed on the substrate LaAlO$_3$ was studied by Matsuno {\it et al.}~\cite{Matsuno}.
Transport and optical properties indicate that this compound has either a very small Mott insulating gap or semiconducting property with rapidly increasing resistivity with decreasing temperature.  The gap amplitude is nearly zero and it can be easily metallized by Sr doping~\cite{Matsuno}.
The magnetic susceptibility appears to show presumable antiferromagnetic transition at around 50K.

\begin{figure}
\includegraphics[width=8cm]{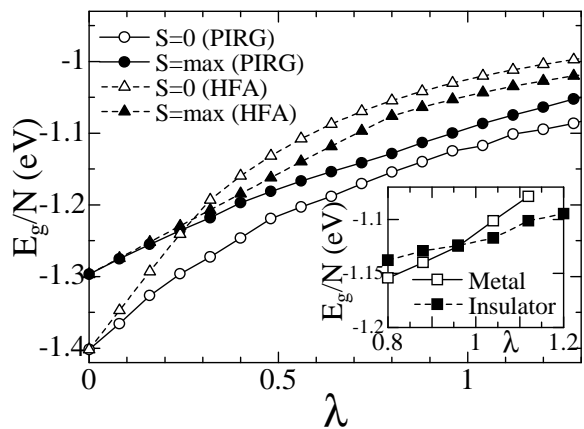}
\caption{Lowest energies per unit cell of total S=0 (open symbols) and ferromagnetic states (filled symbols) by quantum-number projected Hartree-Fock (triangles) and PIRG (circles) calculations for the downfolded model of Sr$_2$VO$_4$.  The energy unit is eV. (Inset):Lowest energies per unit cell of metallic (open squares) and insulating states (filled squares) by quantum-number projected PIRG calculations for the downfolded model of Sr$_2$VO$_4$.}
\label{energy.eps}
\end{figure}

In sharp contrast to this nearly insulating transport properties, the LDA calculation predicts a good metallic behavior (Fig.\ref{Fig.SVOLDADispersion}).  
On the other hand, the Hartree Fock calculation predicts a clear ferromagnetic insulating phase at the realistic parameter values (see Fig.~\ref{energy.eps}). 

\begin{figure}
\includegraphics[width=8cm]{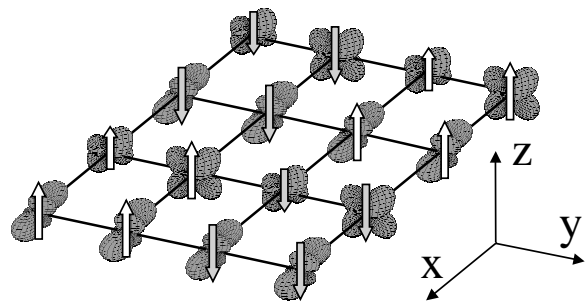}
\caption{Ordered spin-orbital patterns in plane for Sr$_2$VO$_4$ in the Mott insulating ground state with the unit cell of 2 $\times$ 4. Ordered spin moment is proportional to the length of arrows.}
\label{orb_order2.eps}
\end{figure}

The PIRG results indicate that after quantum fluctuations taken into account, the ground state becomes close to the metal-insulator transition point and becomes total spin $S=0$ in contrast to the Hartree-Fock prediction as we see in Fig.~\ref{energy.eps}.
The Mott transition occurs at $\lambda\sim0.95$ as indicated by the jump of the double occupancy $D$ and the realistic parameter is on the verge of the first-order Mott transition. The double occupancy is defined by $D=\langle \sum_{m\sigma} n_{im\sigma}\rangle$. This is consistent with the experimental result.
Furthermore, the spin and orbital are strongly coupled each other and the PIRG results in the spin-orbital order shown in Fig.~\ref{orb_order2.eps} with a remarkable and nontrivial orbital-stripe structure for the insulating phase, while it remains paramagnetic in the metallic side. The realistic value of the relativistic spin-orbit coupling does not seem to alter the present conclusion.  Several possible interlayer configurations are degenerate within the accuracy ($\sim 0.005$eV) of our calculation. At least, the antiferromagnetic state is consistent with the experiments with insulating behavior, while the configurations of spin-orbital order are not experimentally available so far and is desired to test whether this prediction is correct or not. More detailed theoretical study will be reported elsewhere~\cite{Imai-Solovyev-Imada}.

It has turned out that this compound shows very severe competitions.  First, it lies on the verge of the Mott transition.  Second, the ferromagnetic state is rather close in energy to the true ground state with the antiferromagnetic order. Third, candidates of configurations for the spin-orbital order are in severe competitions each other in the order of 100K in energy. The available experimental results are consistent with our present results in contrast to the LDA and Hartree-Fock results. Clearly we need more analyses on this compound.  However, all of the above indicate that our present approach of PIRG combined with the downfolding by using the LDA-GW scheme offers a promising computational method for strongly correlated electron systems.   From the viewpoint of the computational methods, this compound appears to offer a very severe and good benchmark test for the accuracy in taking account of the correlation effects. 

The authors are grateful to F. Aryasetiawan for fruitful discussions and also thank Y. Tokura for clarifications of experimental results on Sr$_2$VO$_4$. 

\end{document}